\documentclass[smallextended]{svjour3}       
\smartqed  
\usepackage{graphicx}
\begin{document}

\title{$\Lambda$ and $\Sigma^+$(1385) hyperons reconstruction in the CBM experiment for p+C reaction at 10 GeV/c.
\thanks{Thanks V. Frise and E. Krishen for kindly response and help with data taking and analysis.}
}


\author{P. Aslanyan  \and
        I. Vassilev
}


\institute{JINR \at Joliot Curie 6, Moscow Region,141980 Dubna, Russia\\
              \email{paslanian@jinr.ru} \\
           \and
           GSI, CBM collaboration \at
              Plunkstrasse 1, Darmstadt
}

\date{Received: date / Accepted: date}

\maketitle

\begin{abstract}
The  efficiency for multi-strange hyperon reconstruction and  the detection strategy for basic kinematic parameters are described using experimental data of p+C interaction at 10 GeV/c. The  efficiency for $\Lambda$ hyperon reconstruction  is equal to 16\%. The $\Sigma^{*+}$(1385)and   $\Sigma^+$(1620)  resonances are observed in the reconstructed  $\Lambda \pi^+$ mass spectrum.

\end{abstract}

\section{Introduction}
\label{intro}

The CBM experiment at FAIR will investigate the
QCD phase diagram at high baryon densities\cite{cbm},\cite{pepan09}.
Energy densities above GeV/$fm^3$ appear at $E_{lab}>$ 5A GeV and exist during the  time interval less than 5 fm/c. Baryon densities above 10$n_0$($n_0$- normal baryon density) can be reached at $E_{lab}>$10A GeV in heavy A+A collisions\cite{mishutin}.
One of the signatures of this new state is the enhanced production of strange particles, therefore hyperon reconstruction is essential for the understanding of the heavy ion collision dynamics. The experimental $\Lambda / \pi^+$ ratio in the $pC$ reaction is approximately two
times larger than this ratio in pp reactions  at the same energy \cite{pepan09}. The study of multi-strange hyperon and exotica production will be one of the major projects of the CBM experiment.

\section{Experimental data}
\label{sec:1}

\subsection{$\Lambda$ identification }
\label{sec:2}

The experimental information of more than 700000 stereo photographs is
used to select events with decays of neutral strange particles (V0 topology). The GEOFIT
based on the GRIND-CERN program is used to measure the momentum vectors associated with the tracks in terms of depth($tg\alpha =y/(\sqrt{(x^2+z^2)})$) and  azimuth angle($\beta =arctg(z/x)$) if  the beam direction is the z axis.
The events with $V^0$ ($\Lambda$ and $K^0_s$)  were identified using the criteria given in \cite{pepan09}. The momentum resolution of $\Lambda$ is equal to 2\%.
The mass of the identified 9838-events with $\Lambda$ hyperon  is consistent with their PDG values.
The total inelastic cross sections for pp and pC interactions at momentum of
10 GeV/c are equal to 30mb and 260 mb\cite{pepan09}, respectively. The contribution from p+p$\to\Lambda$X
 in p+$C_3H_8$ interactions is less than 12\%.

\subsection{Input data}
\label{sec:3}

    These  experimental  data  consist of 9838 preselected events with $\Lambda$- hyperons from $10^6$ inelastic interaction had transformed in the ascii format as an input data  for  CBMROOT framework.
 
 310k central (b = 0) p+C  at 10 GeV events were simulated with UrQMD generator. The following cuts described in \cite{lamcbm} have been used for $\Lambda$ hyperons  reconstruction: $\chi^2_{prim}(p,\pi^-) >$ 3$\sigma$, $\chi^2_{geo}(\Lambda)<$3$\sigma$,  $Z_{vertex} >$ 3cm downstream of the target.

\section{$\Lambda$ and $\Sigma^+$(1385) hyperons reconstruction }
\label{sec:4}

\begin{figure}[ht]
\begin{center}
 {
  \includegraphics[width=25mm,height=30mm]{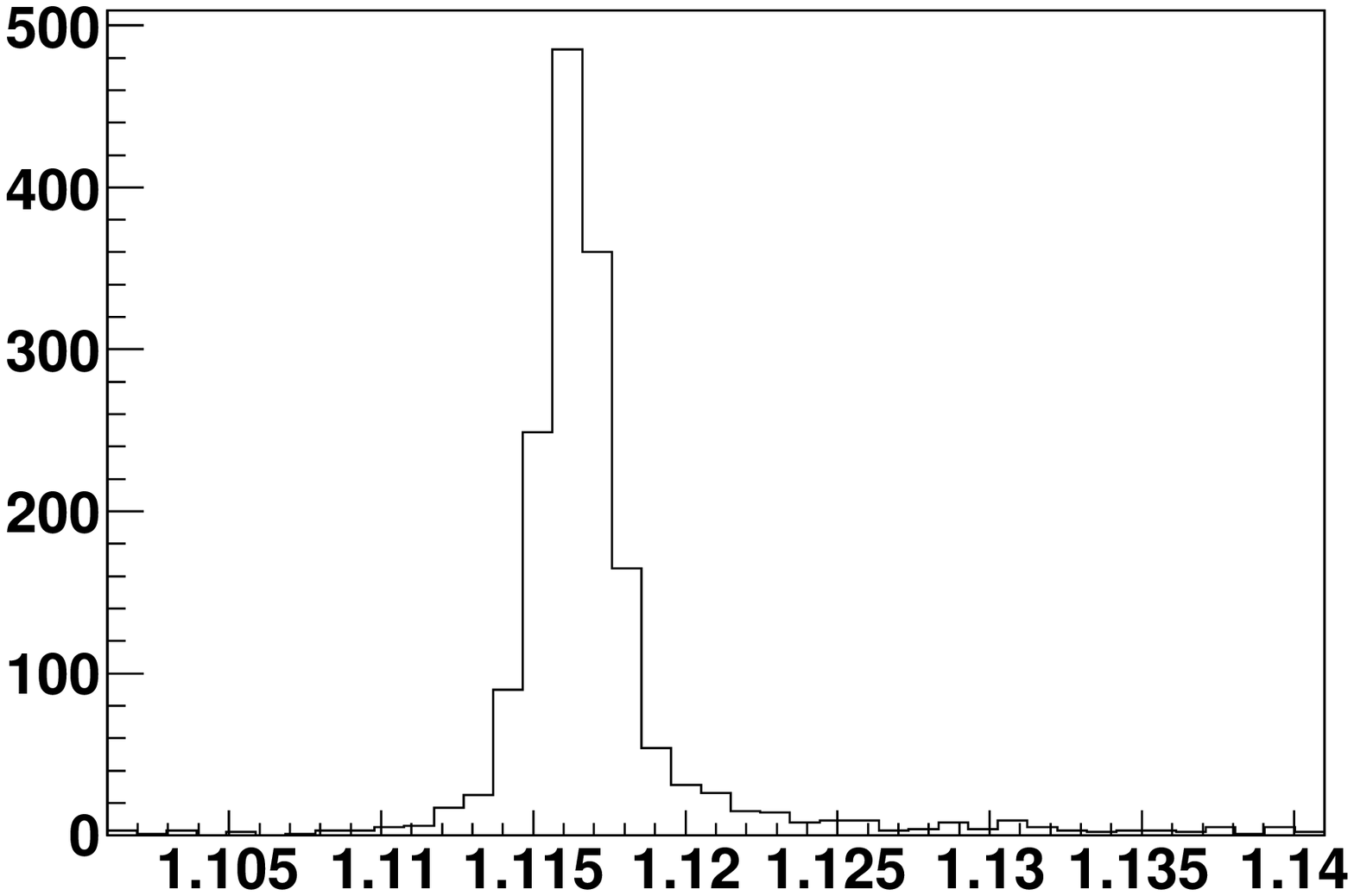}}
  {\large a}
   {
  \includegraphics[width=25mm,height=30mm]{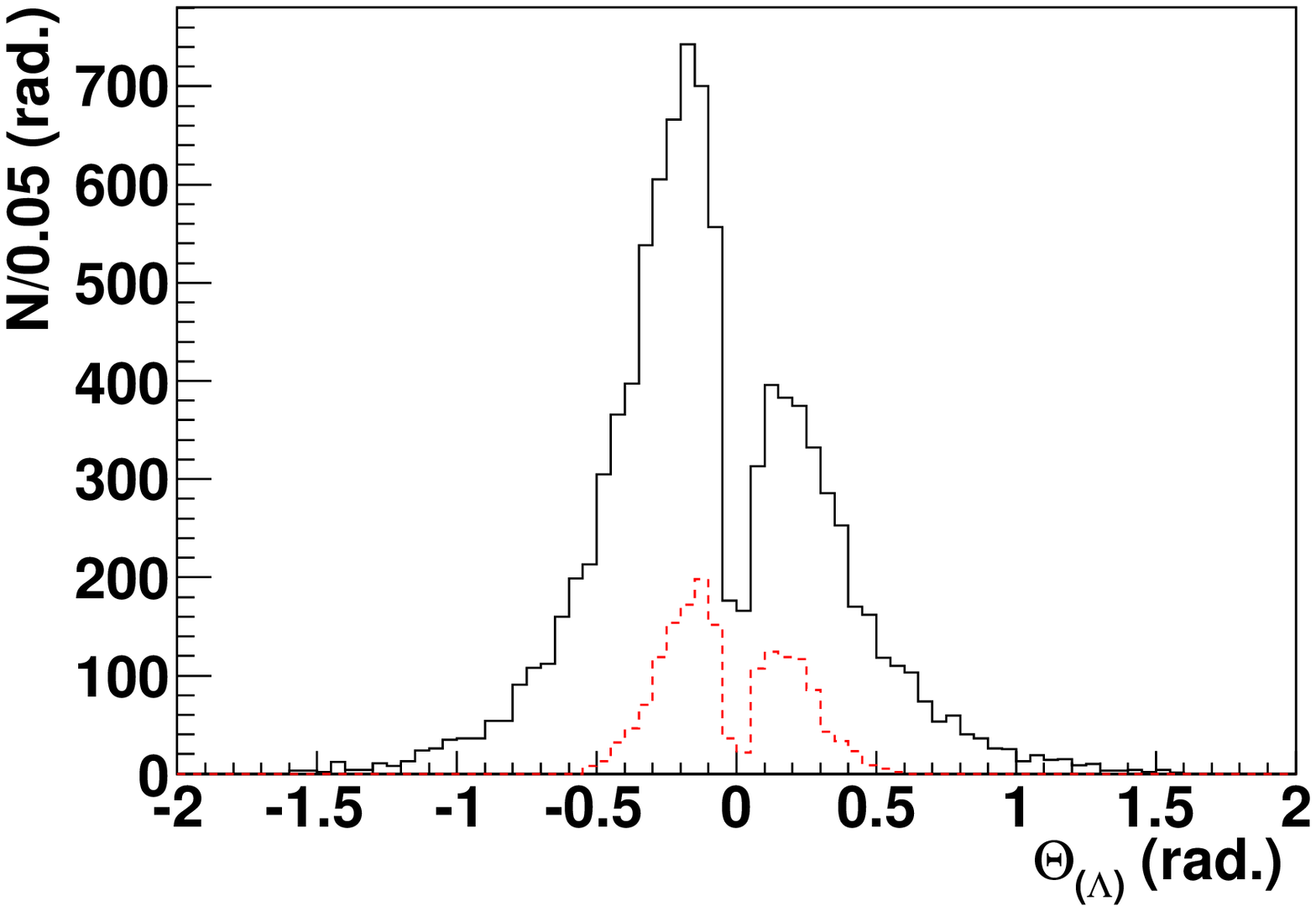}}
   {\large b}
   {
  \includegraphics[width=25mm,height=30mm]{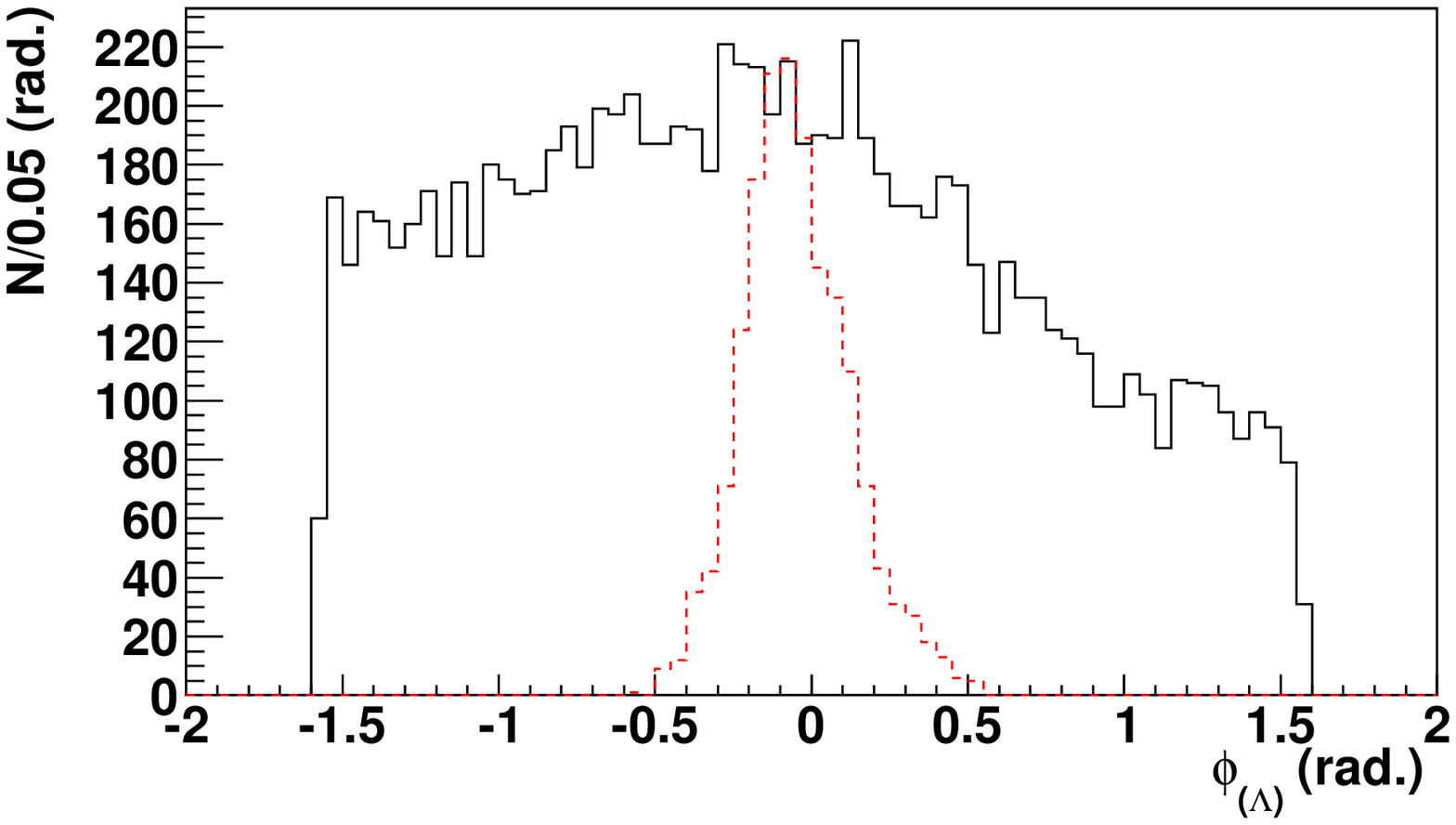}}
   {\large c}
 {
         \includegraphics[width=25mm,height=30mm]{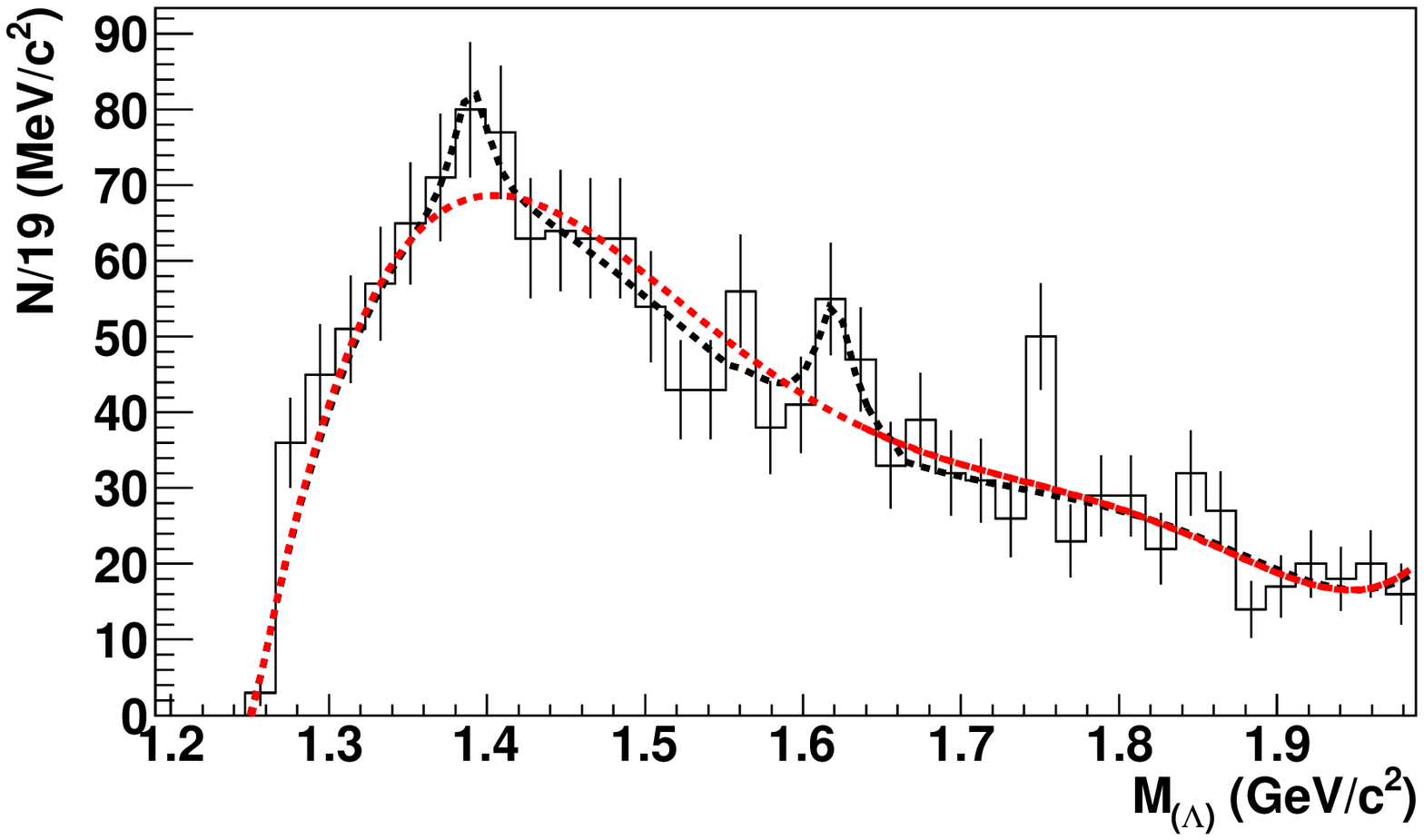}}
   {\large d}
  \caption{The invariant mass distribution for  identified $\Lambda$ hyperons(a).
  The experimental $\Lambda$ hyperons distributions for the scattering angle $\theta$ (b) , azimuth angle $\phi$(c)   and  $\Lambda \pi^+$ invariant mass spectrum  with bin size 19 MeV/$c^2$ (d). The black solid histograms(b,c) are  $\Lambda$ hyperons from all experimental data. The red dashed  histograms are identified $\Lambda$ hyperons from experimental data by CBMROOT.The dashed red curve is the background by the polynomial method. }
\label{lpi}
\end{center}
\end{figure}

   The number of reconstructed experimental and UrQMD events with $\Lambda$ by CBMROOT
   is equal to 1689(without $\chi^2_{geo}$ cut)  and 313 with all selection criteria, respectively.
        The mass and width for identified $\Lambda$ hyperons from experimental data for CBM acceptance is equal
     to $M_{\Lambda}$=1117 and  $\Gamma$=7 MeV/$c^2$, respectively(Fig.\ref{lpi},a).  The evolution of experimental and URQMD events for $\Lambda$ hyperons shows Table \ref{tab:1}.   Fig. \ref{lpi},b,c shows  the  scattering ($\theta <0$, cos$(\theta ) =P_z/P$) and azimuth ($\phi \sim 0$, tg $(\phi)= P_y/P_x$ ) angles distributions in spherical system coordinates for the experimental data with CBM acceptance(the red dashed histogram). P and $P_x, P_y, P_z$ is total momentum and  x,y,z projections, respectively. The beam direction is the z axis.

The curve(Fig.~\ref{lpi},d is the sum of the background by 8 order polynomial and 2 Breit-Wigner function ($\chi^2/N.D.F.=43/38$). The small peaks are observed in mass range of  $\Sigma^+$(1385)and  $\Sigma^+$(1620) with $\approx 3\sigma$.

\section{Conclusion}
\label{sec:5}

The  $\Lambda$ and  $\Sigma^+$(1385) hyperons signals reconstruction efficiency is about  16\% and $\approx$7\%  for  the p+C reaction  at  10 GeV/c (Table \ref{tab:1}), respectively. The obtained results show that the study of hyperons in high statistics experiments with the CBM facility at FAIR  have a high  scientific potential.

\begin{table}
\caption{The evolution of experimental and URQMD events for $\Lambda$ hyperons with acceptance for CBM .
  }
\label{tab:1}       
\begin{tabular}{llllll}
\hline\noalign{\smallskip}
Type of data & The number & The all number & All events with& with criteria\\
 for p+C reaction&  of collisions & of $\Lambda\to\pi^-p$ &$V^0$ topology&$for \Lambda$\\
\noalign{\smallskip}\hline\noalign{\smallskip}
Exp. data&$10^6$&9838&1742&1634\\
\noalign{\smallskip}\hline
URQMD& 3.1$\times 10^5$& 5310(7 mb)&1599& 313 \\
\noalign{\smallskip}\hline
\end{tabular}
\end{table}

\end{document}